\documentclass[12pt]{article}
\begin{document}
\vspace*{-.6in}
\thispagestyle{empty}
\begin{flushright}
CALT-68-2288\\ 
CITUSC/00-043\\
hep-th/0007118
\end{flushright}
\baselineskip = 18pt

\vspace{.5in}
{\LARGE
\begin{center}
String Theory: The Early Years
\end{center}}

\begin{center}
John H. Schwarz\footnote{Work supported in
part by the U.S. Dept. of Energy under Grant No.
DE-FG03-92-ER40701.}
\\
\emph{California Institute of Technology\\
Pasadena, CA  91125, USA}
\end{center}
\vspace{1in}

\begin{center}
\textbf{Abstract}
\end{center}
\begin{quotation}
\noindent
Lenny Susskind has made many important contributions to theoretical physics
during the past 35 years.
 
In this talk I will discuss the early history of string theory
(1968-72)  emphasizing Susskind's contributions.
\end{quotation}

\vfil
\centerline{\it Presented at Symposium for L. Susskind's 60th Birthday}

\newpage

\pagenumbering{arabic}

\section{S-Matrix Theory, Duality, and the Bootstrap}

In the late 1960s there were two parallel trends in particle
physics. On the one hand, many hadron resonances were discovered,
making it quite clear that hadrons are not elementary particles.
In fact, they were found, to good approximation, to lie on linear
parallel Regge trajectories, which supported the notion that they
are composite. Moreover, high energy scattering data displayed
Regge asymptotic behavior that could be explained by the
extrapolation of the same Regge trajectories, as well as one with
vacuum quantum numbers called the {\it Pomeron}. This set of
developments was the focus of the S-Matrix Theory community of
theorists. The intellectual leader of this community was Geoffrey
Chew at UC Berkeley. One popular idea espoused by Chew and
followers was ``nuclear democracy" -- that all hadrons can be
regarded as being equally fundamental. A more specific idea was
the ``bootstrap", that the forces arising from hadron exchanges
are responsible for binding the hadrons, as composites of one
another, in a more or less unique self-consistent manner.

The second major trend in the late 1960s grew out of the famous
SLAC experiments on deep inelastic electron scattering. These gave
clear evidence for point-like constituents (quarks and gluons)
inside the proton. This led to Feynman's ``parton'' model, which
was also an active area of research in those days. I would submit
that Susskind was philosophically more in tune with the parton
world view than the S matrix one, though he clearly had a foot in
each camp. My Berkeley training put me solidly in the
S-matrix camp.

In the early 1970s, it became clear that QCD is the correct theory
of strong interactions. It can be used for explicit computations in
large momentum regimes where perturbation
theory can be used, thanks to asymptotic freedom. Modern
phenomenological studies in such regimes are directly descended
from the older parton models. In other regimes, such as
diffraction scattering, where perturbation theory is not
applicable, S-matrix ideas are still used. So both major trends
from the late 1960s have their descendents 30 years later.

String theory, which is the subject I want to focus on here, grew
out of the S-Matrix approach to hadronic physics. The bootstrap
idea got fleshed out in the late 1960s with the  notion of a
duality relating $s$-channel and $t$-channel processes that went
by the name of ``finite energy sum rules'' \cite{Dolen:1967} -
\cite{Harari:1968}. Another influential development was the
introduction of ``duality diagrams'', which keep track of how
quark quantum numbers flow in various processes \cite{Harari:1969,
Rosner:1969}. Susskind contributed papers on this topic
\cite{Susskind:1970a, Susskind:1970b}. Later, duality diagrams
would be reinterpreted as string world-sheets, with the quark
lines defining boundaries. A related development that aroused
considerable interest was the observation that the bootstrap idea
requires a density of states that increases exponentially with
mass, and that this implies the existence of a critical
temperature, called the Hagedorn temperature \cite{Hagedorn:1968}
- \cite{Frautschi:1971}. It was generally supposed that this it is
an ultimate temperature, though a phase transition was clearly
another possibility.

\section{The Dual Resonance Model}

The duality program got a real shot in the arm in 1968 when
Veneziano found a specific mathematical function that explicitly
exhibits the features that people had been discussing in the
abstract \cite{Veneziano:1968}. The function, an Euler beta
function, was proposed to give a good phenomenological description
of the reaction $\pi + \omega \to \pi + \pi$ in the narrow
resonance approximation. This was known to be a good
approximation, because near linearity of Regge trajectories
implies that the poles should be close to the real axis. A little
later Lovelace and Shapiro proposed a similar formula to describe
the reaction $\pi + \pi \to \pi + \pi$ \cite{Lovelace:1968,
Shapiro:1969}. Chan and Paton explained how to incorporate
``isospin'' quantum numbers in accord with the Harari--Rosner
rules \cite{Paton:1969}. Also, within a matter of months Virasoro
found an alternative formula with many of the same duality and
Regge properties that required full $s$-$t$-$u$ symmetry
\cite{Virasoro:1969a}. Later it would be understood that whereas
Veneziano's formula describes scattering of open-string ground
states, Virasoro's describes scattering of closed-string ground
states.

In 1969 several groups independently discovered $N$-particle
generalizations of the Veneziano four-particle amplitude
\cite{Bardakci:1969a} - \cite{Koba:1969b}. The $N$-point
generalization of Virasoro's four-point amplitude was constructed
by Shapiro \cite{Shapiro:1970}. In short order Fubini and
Veneziano, and also Bardakci and Mandelstam, showed that the
Veneziano $N$-particle amplitudes could be consistently factorized
in terms of a spectrum of single-particle states described by an
infinite collection of harmonic oscillators \cite{Fubini:1969a} -
\cite{Fubini:1970}. This was a striking development, because it
suggested that these formulas could be viewed as more than just 
approximate phenomenological descriptions of hadronic scattering.
Rather, they could be regarded as the tree approximation to a
full-fledged quantum theory. I don't think that anyone had
anticipated such a possibility one year earlier. It certainly came
as a surprise to me.

One problem that was immediately apparent was that since the
oscillators transform as Lorentz vectors, the time components
would give rise to negative-norm ghost states. Everyone knew that
such states would violate unitarity and causality. Virasoro came
to the rescue by identifying an infinite set of subsidiary
conditions, which plausibly could eliminate the negative-norm
states from the spectrum \cite{Virasoro:1970, DD:1970}. These subsidiary
conditions are defined by a set of operators, which form the
famous Virasoro algebra \cite{Fubini:1971}. The central term in the
algebra was discovered by Joe Weis (unpublished). One price for
eliminating ghosts in the way suggested by Virasoro was that the
leading open-string Regge trajectory had to have unit intercept,
and hence, in addition to a massless vector, it contributes a
tachyonic ground state to the spectrum \cite{Fairlie}.

\section{The String Idea}

Once it was clear that we were dealing with a system with a rich
spectrum of internal excitations, and not just a bunch of
phenomenological formulas, it was natural to ask for a physical
interpretation. The history of who did what and when is a little
tricky to sort out.  As best I can tell, the right answer -- a
one-dimensional extended object (or ``string'') -- was discovered
independently by three people: Nambu, Susskind, and Nielsen.
Nambu's contribution to the 1969 conference held at Wayne State
University apparently was first, but the conference was rather
obscure
 \cite{Nambu:1970a}, and the paper was not widely circulated.
His subsequent paper, submitted to a symposium in Copenhagen,
proposed that the string action is the area of the world-sheet in
analogy with the proper length of the world-line of a point
particle \cite{Nambu:1970b}.  These difficult to find papers have been
reprinted in \cite{Nambu:1995}.

The string idea also appears quite
clearly in Susskind papers, which were published in a refereed
journal \cite{Susskind69, Susskind:1970e, Susskind:1970f}. Nielsen's first
paper on the subject was submitted to a 1970 conference in Kiev,
though it was not published in the proceedings
\cite{Nielsen:1970a}. However, shortly thereafter, he and Fairlie
described their approach in a refereed journal
\cite{Fairlie:1970}. The Nielsen papers emphasize an analog
electrostatic model in which one solves Laplace's equation on a
disk with sources on the boundary. This is just the string wave
equation on a Euclideanized world sheet.  The electrostatic
analogy was also discussed by Shapiro for the Virasoro--Shapiro
model taking the domain to be a sphere rather than a disk and with
the sources attached to the interior of the surface
\cite{Shapiro:1970}. We recognize this to be the proper
description of tree-level closed-string amplitudes.

A somewhat different approach treated the string world-sheet as
some kind of sum or limit of complicated planar Feynman diagrams,
sometimes referred to as fishnet diagrams. This approach, along
with various related parton ideas, was pursued by Sakita and
Virasoro \cite{Sakita:1970a} as well as by Susskind and Nielsen
\cite{Nielsen:1970b} - \cite{Susskind:1972b}. Note that these
works are several years prior to `t Hooft's famous paper on
large-N gauge theory \cite{tHooft:1974}. They were not very
specific, however, about which field theory should be used to form
the fishnet diagrams or what limit was required to make contact
with a string world sheet. 

Though there was some progress in the intervening years \cite{MansouriA, MansouriB},
the string interpretation of the dual resonance model was not very
influential in the development of the subject until the appearance
of the 1973 paper by Goddard, Goldstone, Rebbi, and Thorn
\cite{Goddard:1973}. It explained in detail how the string action
could be quantized in light-cone gauge. Subsequently Mandelstam
extended this approach to the interacting theory
\cite{Mandelstam:1973}.

\section{Loop Amplitudes}

From the factorization of $N$-particle amplitudes one had learned
the tree approximation spectrum and couplings. With this
information in hand, it became possible to construct one-loop
amplitudes. The first such attempt was made by Kikkawa, Sakita,
and Virasoro \cite{Kikkawa:1969}. 
They did not have enough information in hand to do
it completely right, but they pioneered many of the key ideas and
pointed the way for their successors \cite{Kikkawa:1970} - \cite{Frampton:1971}. 

Those who worked on
this problem included a group at Princeton consisting of David
Gross, Andr\'e Neveu, Jo\"el Scherk, and myself.
Another group consisted of Lenny Susskind and various collaborators.
Neveu and Scherk  studied the divergence of the one-loop planar amplitude.
They realized that by performing a Jacobi transformation of the theta
functions in the integrand they could isolate the divergent piece and propose
a fairly natural counterterm \cite{Neveu:1970}. 
Essentially the same thing was done independently
by Susskind and Frye \cite{Frye:1970a}. The modern interpretation of these
results is that viewed in a dual channel there is a closed string going into the vacuum.
The divergence can be attributed to the tachyon in that channel. Its contribution
is the piece they subtracted. This also explains why in a model without tachyons
such divergences would not occur. The cancellation of the milder divergences due
to dilaton tadpoles also became an important consideration in later years.

One of the important things discovered by the Princeton group \cite{Gross:1970b},
which was discovered independently by Frye and Susskind at about
the same time \cite{Frye:1970b}, was that the nonplanar loop amplitude gives new and
unexpected singularities. In both of these works the dimension of
spacetime was assumed to be four, and the Virasoro subsidiary
constraints were not implemented on the internal states
circulating in the loop. As a result the singularities were found
to be unitarity-violating branch points.

Lovelace observed that if one allows the
spacetime dimension to be 26 and supposes that the subsidiary
conditions imply that only transverse oscillators contribute, then
instead of branch points the singularities would be poles \cite{Lovelace:1971}. 
As we now know, these are are
the closed-string poles in the nonplanar open-string loop. This
calculation showed that unitarity requires that one choose the
critical dimension and the intercept value for which the Virasoro
conditions are satisfied. In those days these states were referred
to as Pomeron-states rather than closed string states, since they
necessarily carried vacuum quantum numbers. It was supposed that
the unphysical Regge intercept value of two -- implying the
existence of a massless spin two particle as well as a tachyon --
would somehow be lowered to the desired value of one in a more
realistic model.

Shapiro carried out the first closed-string loop calculation \cite{Shapiro:1972}.
Specifically, he computed the one-loop (torus) amplitude.

\section{Conclusion}

The younger generation of theoretical physicists is aware of the wide range
of important contributions that Lenny Susskind has made to our field
in recent years. What they may be less aware of is his
important role in the early history of string theory. I hope this presentation will
help to publicize that work.

\end{document}